\documentclass[conference]{IEEEtran}
\usepackage{subfigure} 
\usepackage{bm} 
\usepackage{graphicx,graphics} 
\usepackage{amsmath, amsthm, amsfonts, amssymb, amsbsy}
\usepackage{setspace} 
\usepackage{pstricks,arydshln}
\usepackage{multirow,multicol} 
\usepackage{booktabs}
\usepackage{mathbbol,algorithmic,amscd,textcomp,amssymb,epsfig,graphics,epsf,color,amsmath,balance,algorithm,cite}
\usepackage{caption} 
\usepackage{cite} 
\usepackage{tabularx}
\usepackage{makecell}
\usepackage{epstopdf}

\begin{document}

\title{Radio-Frequency Multi-Mode OAM Detection Based on UCA Samples Learning}

\author{Jiabei Fan$^1$, Rui Chen$^{1,2}$, Wen-Xuan Long$^{1,3}$, Marco Moretti$^3$, and Jiandong Li$^1$\\
$^1$State Key Laboratory of Integrated Service Networks (ISN), Xidian University, Shaanxi, China\\
$^2$National Mobile Communication Research Laboratory, Southeast University, Nanjing, China\\
$^3$University of Pisa, Dipartimento di Ingegneria dell'Informazione, Italy\\
Email: jiabeifan@stu.xidian.edu.cn, rchen@xidian.edu.cn

}

\maketitle

\begin{abstract}
Orbital angular momentum (OAM) at radio-frequency provides a novel approach of multiplexing a set of orthogonal modes on the same frequency channel to achieve high spectral efficiencies. However, classical phase gradient-based OAM mode detection methods require perfect alignment of transmit and receive antennas, which greatly challenges the practical application of OAM communications. In this paper, we first show the effect of non-parallel misalignment on the OAM phase structure, and then propose the OAM mode detection method based on uniform circular array (UCA) samples learning for the more general alignment or non-parallel misalignment case. Specifically, we applied three classifiers: K-nearest neighbor (KNN), support vector machine (SVM), and back-propagation neural network (BPNN) to both single-mode and multi-mode OAM detection. The simulation results validate that the proposed learning-based OAM mode detection methods are robust to misalignment errors and especially BPNN classifier has the best generalization performance.
\end{abstract}

\vspace{-0mm}
\section{Introduction}
\vspace{-0mm}

Since the famous `Venice experiment' in 2012 \cite{Tamburini2012Encoding} showing that orbital angular momentum (OAM) can be succesfully applied to radio-frequency (RF) transmissions, a significant research effort has been invested into the OAM wireless communications as a novel approach for multiplexing a set of orthogonal modes on the same frequency channel and achieving high spectral efficiencies \cite{Chen2020Orbital,Mahmouli20134,Yan2014High,Zhang2017Mode,
Ren2017Line,Chen2018Beam,Chen2018A,Yagi2019200,Chen2020Multi-mode,Chen2020Generation,Long2021AoA}. So far, there are a number of approaches to generate radio OAM beams \cite{Yan2014High,Liang2016Orbital,Zhang2016Generation,Chen2020Generation,
Chen2018A,Chen2020Multi-mode,Zhang2017Four,Yu2016Generating}, among which uniform circular array (UCA) is the most popular way because of the ubiquitous application of antenna array in 4G, 5G and even 6G base stations (BSs) and its compatibility with MIMO beam steering \cite{Chen2018Beam}. The highest data rate 200 Gbps of OAM communication system has been realized with concentric UCAs by NTT\cite{Yagi2019200}. Therefore, we adopted UCA instead of other particular antenna structures for OAM generation and reception in our study.

The successful signal transmission with no matter OAM multiplexing or index modulation depends on the precondition of correctly detecting the received OAM modes. In \cite{Mohammadi2010Orbital,Allen2014Wireless}, the RF single-mode OAM detection is proposed for the first time and solved by using single point method and phase gradient method, respectively. Then, the RF dual-mode OAM beam measurement method is investigated in \cite{Xie2017Mode}. Moreover, the experiment of digitally rotating a virtual antenna to detect OAM mode based on the relationship between the rotational Doppler frequency shift and OAM mode is implemented in \cite{Zhang2017Detecting}. In \cite{Chen2018Detection}, multiple OAM beams are detected with a metasurface through converting an OAM beam into a Gaussian one, and then determining the incident OAM mode by locating.

The above existing multi-mode OAM detection methods require accurate alignment of the transmit and receive antennas, otherwise the methods fail to work or the detection performance seriously deteriorates. However, the requirement of accurate alignment is difficult to meet in practical environments \cite{Chen2018Beam}. Fortunately, machine learning (ML), which is commonly used to solve complex pattern recognition problems, has been applied to optical OAM mode detection to extract the features of the intensity distributions \cite{Xiong2020Convolutional,Wang2019Efficient}. Since the wireless OAM communication receiver can only leverage the UCA to receive limited number of spatial sampling signals, the convolutional neural network (CNN)-based image detection method used for optical OAM detection cannot be directly applied to the RF OAM mode detection.
Therefore, in this paper we propose three OAM mode detection methods based on K-nearest neighbor (KNN), support vector machine (SVM), and back-propagation neural network (BPNN) classifiers with array signal phase information. The simulation results validate that the proposed multi-mode OAM detection methods are efficient and robust in the non-parallel misalignment cases.

\section{System Model}
The harsh precondition for the perfect alignment of the transmit and receive antennas of the OAM communication system is difficult to keep all the time in practice, so we consider the basic non-parallel case \cite{Chen2018Beam} as shown in Fig. \ref{Fig1}.

\vspace{-0mm}
\subsection{Channel Model}
\begin{figure}[t] 
\setlength{\abovecaptionskip}{0cm}   
\setlength{\belowcaptionskip}{-0.4cm}   
\begin{center}
\includegraphics[scale=0.3]{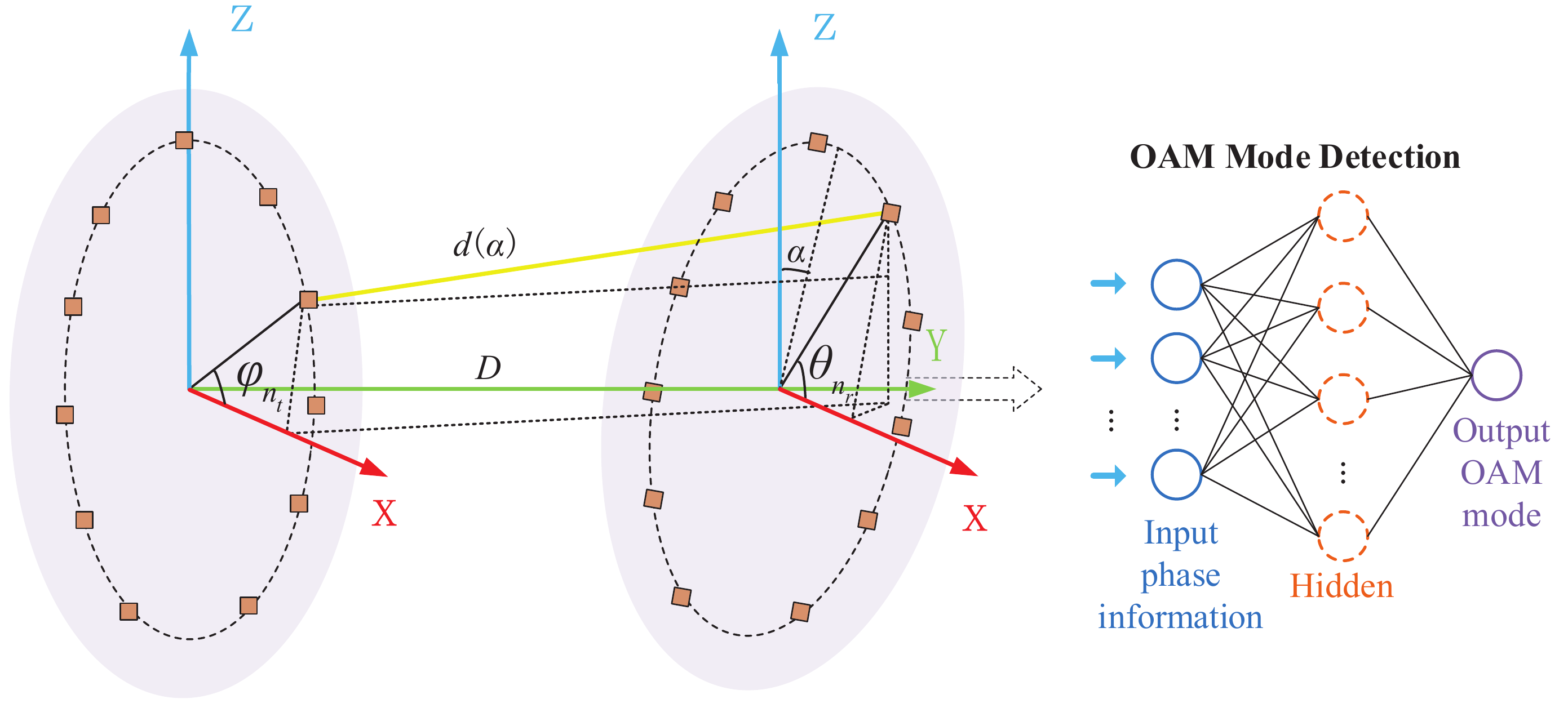}   
\end{center}
\caption{The OAM mode detection of the UCA-based non-parallel misalignment system.}
\label{Fig1}
\end{figure}

In line-of-sight (LoS) communications, propagation through the RF channel leads to attenuation and phase rotation of the transmitted signal. This effect is modelled by the multiplication for a complex constant $h$, whose value  depends on the distance $d$ between the transmit and the receive antennas:
\begin{align} \label{FreeSpaceChannel}
h(d) = \beta \frac{\lambda }{{4\pi d}}\exp \left( - j\frac{{2\pi d}}{\lambda }\right),
\end{align}
where $\lambda$ is the wavelength, and $\displaystyle \lambda/4\pi d$ denotes the degradation of amplitude, and the complex exponential term is the phase difference due to the propagation distance. The term $\beta$ models all constants relative to the antenna elements and their patterns. Thus, according to (\ref{FreeSpaceChannel}) and the geometric relationship depicted in Fig. \ref{Fig1}, the channel coefficients from the $n_t$th ($1\le n_t \le N_t$) transmit antenna element to the $n_r$th ($1\le n_r \le N_r$) receive  antenna element can be expressed as $h_{n_r,n_t}=h(d_{n_r,n_t})$, where the transmission distance $d_{n_r,n_t}$ is calculated as
\begin{align} \label{dmn}
&d_{n_r,n_t}=[R_r^2+R_t^2+D^2+2DR_r\sin{\theta _{n_r}}\sin\alpha\nonumber\\
&-2R_rR_t(\cos{\varphi _{n_t}}\cos{\theta _{n_r}}+\sin{\varphi _{n_t}}\sin{\theta _{n_r}}\cos\alpha)]^{1/2},
\end{align}
where $D$ is the distance between the transmit and the receive UCA centers, $R_t$ and $R_r$ are respectively the radii of the transmit and the receive UCAs, $\varphi=[2\pi(n_t-1)/N_t+\varphi_0]$ and $\theta=[2\pi(n_r-1)/N_r+\theta_0]$ are respectively the azimuthal angles of the transmit and the receive UCAs, $\varphi_0$ and $\theta_0$ are the corresponding initial angles of the first reference antenna element in both UCAs, $\alpha$ is the oblique angle shown in Fig. \ref{Fig1}. Thus, the channel matrix of the UCA-based LoS OAM communication system can be expressed as $\mathbf{H}=[h(d_{n_r,n_t})]_{N_r\times N_t}$. When $\alpha = 0$, $\mathbf{H}$ is a circulant matrix that can be decomposed by $N$-dimensional discrete Fourier transform (DFT) matrix \cite{Chen2018Beam}.

\subsection{Signal Model}
As the UCA can generate multi-mode OAM beams with the baseband (partial) DFT matrix $\mathbf{F}_U$ \cite{Chen2020Multi-mode}, the equivalent baseband signal models of multi-mode OAM symbols sent by the transmit UCA can be expressed as $\mathbf{F}_U^H\mathbf{s}$, where $\mathbf{s}=[s(\ell_1),s(\ell_2),\cdots,s(\ell_U)]^T$ is the modulation symbol vector transmitted on $U$ OAM modes, $\mathbf{F}_U=[\mathbf{f}^H(\ell_1),$ $\mathbf{f}^H(\ell_2),\cdots,\mathbf{f}^H(\ell_U)]^H$, and $\mathbf{f}(\ell_u)$ $=$ $\big[f_{1_t}(\ell_u),$ $f_{2_t}(\ell_u),$ $\cdots,$ $f_{N_t}(\ell_u)\big]$ $=$ $\frac{1}{\sqrt{N_t}}[1,e^{-i\frac{2\pi \ell_u}{N_t}},\cdots,$ $e^{-i\frac{2\pi \ell_u(N_t-1)}{N_t}}]$. For easier analysis, we assume $\mathbf{s}=\mathbf{1}_U$ in the training stage, and $\mathbf{1}_U$ is $U$-dimensional column vector of ones.
Then, the received baseband symbol vector can be written as:
\begin{align} \label{x_total}
{\bf{x}}=\mathbf{H}\mathbf{F}_U^H\mathbf{1}_U+\mathbf{z},
\end{align}
where ${\bf{x}}=[x_1,x_2,\cdots,x_{N_r}]^T$, and $x_{n_r}$ can be expressed as
\begin{align}\label{xnr}
x_{n_r}&=\mathbf{h}_{n_r}\mathbf{f}_t+z_{n_r} \nonumber\\
&=\frac{\beta}{2k}\sum_{n=1}^{N_t} \frac{e^{ik|\bm{d}_{{n_r},{n_t}}|}}{|\bm{d}_{{n_r},{n_t}}|}\sum_{u=1}^{U}e^{i\ell_{u}\varphi_n}
+z_{n_r}\nonumber\\
&=\frac{\beta}{2k}\sum_{n_t=1}^{N_t} \frac{e^{ik|\bm{D}-\bm{r}'_{n_t}+\bm{r}'_{n_r}|}} {|\bm{D}-\bm{r}'_{n_t}+\bm{r}'_{n_r}|}\sum_{u=1}^{U}e^{i\ell_{u}\varphi_{n_t}}
+z_{n_r}\nonumber\\
&\approx\frac{\beta}{2k}e^{i\bm{k}\cdot\bm{r}_{n_r}}\bigg[\frac{e^{ikD}}{D}
\sum_{n_t=1}^{N_t} e^{-i\bm{k}\cdot\bm{r}_{n_t}}\sum_{u=1}^{U}e^{\ell_{u}\varphi_{n_t}}\bigg] +z_{n_r},
\end{align}
$\mathbf{h}_{n_r}$ $=$ $[h_{{n_r},1},$ $h_{{n_r},2},$ $\cdots,$ $h_{{n_r},N_t}]$, $\mathbf{f}_t$ $=$ $\big[\sum_{u=1}^{U}f_{1_t}(\ell_u),$ $\sum_{u=1}^{U}f_{2_t}(\ell_u),$ $\cdots,$ $\sum_{u=1}^{U}f_{N_t}(\ell_u)\big]^T$, $\bm{d}_{{n_r},{n_t}}$ is the position vector from the $n_t$th transmit antenna element to the $n_r$th receive antenna element, $i = \sqrt { - 1}$ is the imaginary unit, $\bm{k}$ is the wave vector. The far-field approximations are $|\bm{D}-\bm{r}'_{n_t}+\bm{r}'_{n_r}| \approx D$ for amplitudes and $|\bm{D}-\bm{r}'_{n_t}+\bm{r}'_{n_r}| \approx D - \bm{\hat{D}} \cdot \bm{r}_{n_t} + \bm{\hat{D}} \cdot \bm{r}_{n_r}$ for phases, where $\bm{\hat{D}}$ is the unit vector of $\bm{D}$, $\bm{r}_{n_t}$ and $\bm{r}_{n_r}$ are the position vectors of transmit and receive antenna elements respectively, ${\bf{z}}=[z_1,z_2,\cdots,z_{N_r}]^T$ is noise vector.

In the offline training stage, the phases of the spatial samples at the receive UCA are obtained according to the received signal model \eqref{xnr} and then collected in the data set $\rm{S}$, i.e., $\rm{S}$ $=$ $\{(\text{arg}({\bf{\bar{x}}}(D_p,\alpha_q)),{\ell}^{p,q})|p=1,2,\cdots,P; q=1,2,\cdots,Q\}$, where $D_p$ is the $p$th item of $P$ distances, $\alpha_q$ is the $q$th item of $Q$ oblique angles, $\text{arg}(\cdot)$ denotes the argument of a complex number, ${\bf{\bar{x}}}(D_p,\alpha_q)$ is the noiseless sample corresponding to the position $(D_p,\alpha_q)$ and can be expressed as
\begin{align} \label{Feature_x_multi}
\text{arg}({\bf{\bar{x}}}(D_p,\alpha_q)) &\!=\!\text{arg}\left(\!e^{ikD} e^{i\bm{k}\cdot\bm{r}_{n_r}}\sum_{n_t=1}^{N_t} e^{-i\bm{k}\cdot\bm{r}_{n_t}}\sum_{u=1}^{U}e^{\ell_{u}\varphi_{n_t}}\!\right)\nonumber \\
&\!=\! \text{arg} \left(\sum_{n_t=1}^{N_t}h(d_{{n_r},{n_t}})\sum_{u=1}^{U}f_{n_t}(\ell_u)\right),
\end{align}
and ${\ell}^{p,q}$ denotes the label corresponding to $\text{arg}({\bf{\bar{x}}}(D_p,\alpha_q))$.

\section{Single-mode And Multi-mode OAM Detection Method based on Samples Learning}
In this section, we propose three different OAM detection methods to realize OAM single-mode and multi-mode detection in more general non-parallel misalignment case.

\subsection{Problem Description}
\begin{figure}[t]  
\setlength{\abovecaptionskip}{0cm}   
\setlength{\belowcaptionskip}{-0.0cm}   
\centering
\subfigure[]{\includegraphics[scale=0.3]{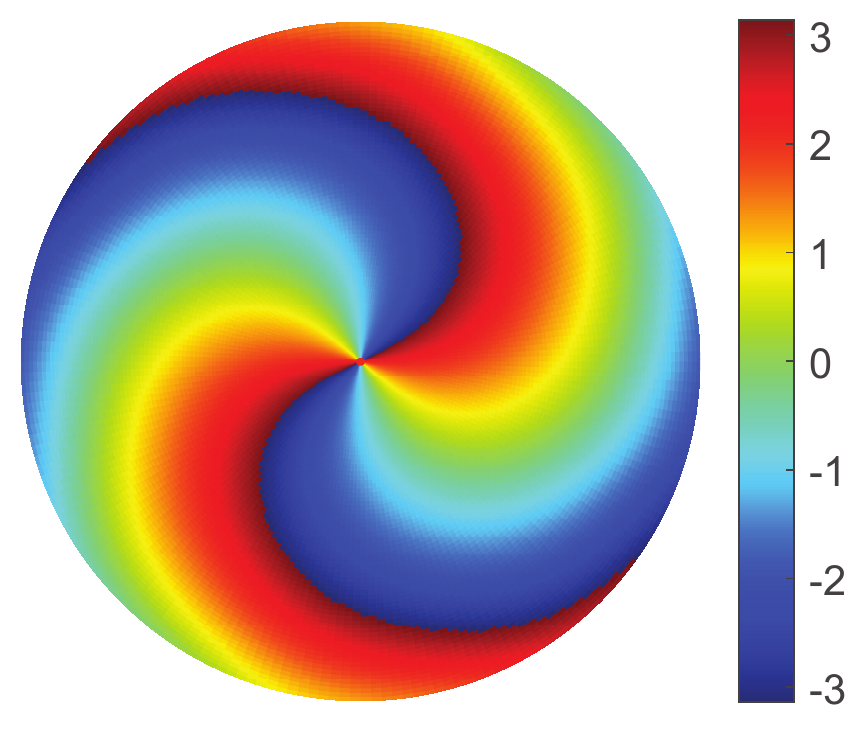}}\hspace{5.7mm}
\quad
\subfigure[]{\includegraphics[scale=0.3]{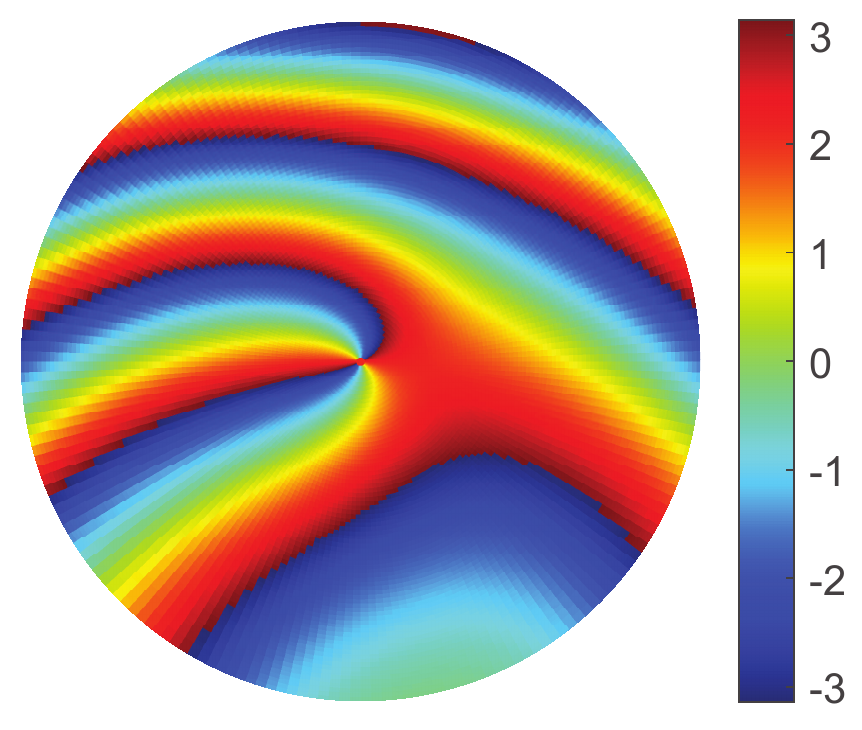}}
\quad
\subfigure[]{\includegraphics[scale=0.3]{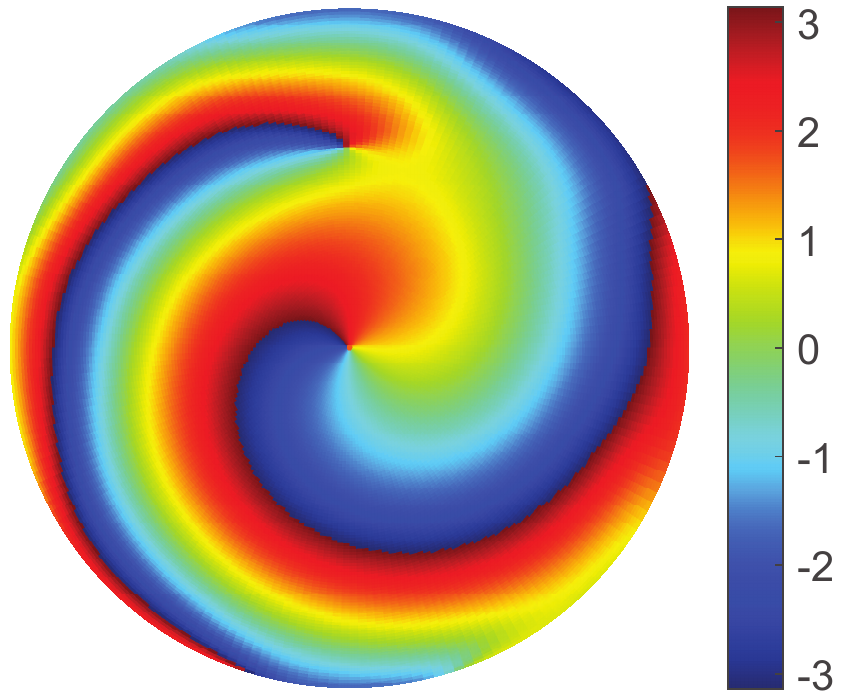}}\hspace{6.5mm}
\quad
\subfigure[]{\includegraphics[scale=0.3]{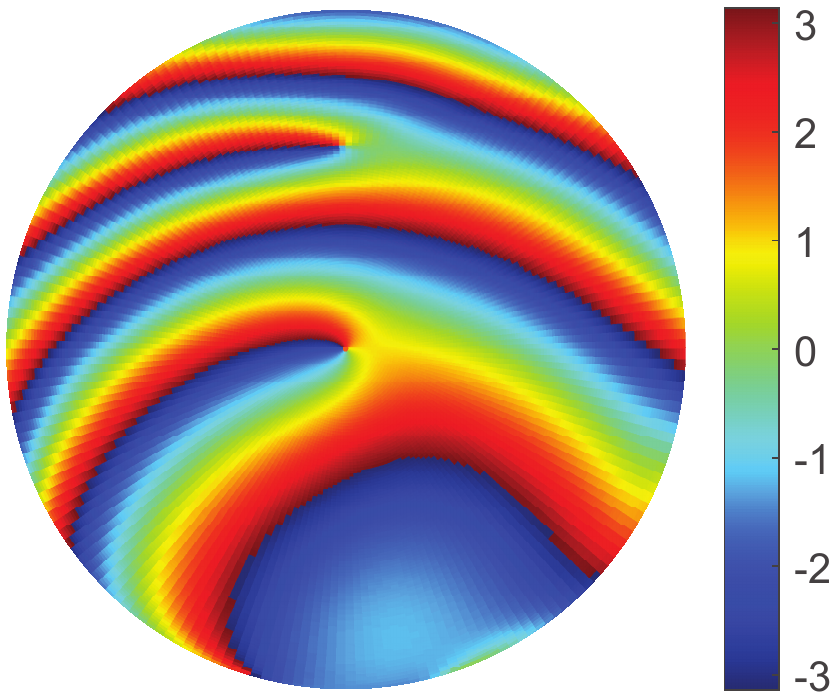}}
\caption{
The rotational phase of single-mode OAM and multi-mode OAM beams simulated by MATLAB in the perfect alignment and misalignment cases: (a)$\ell=+2$, $\alpha=0^{\circ}$, (b)$\ell=+2$, $\alpha=5^{\circ}$, (c)$\ell_1=+1$, $\ell_2=+2$ and $\alpha=0^{\circ}$, (d) $\ell_1=+1$, $\ell_2=+2$ and $\alpha=5^{\circ}$.}
\label{Fig2}
\end{figure}

It is known that the classical OAM detection methods are based on the circular phase gradient. No matter single-mode or dual-mode detection, circular phase gradient methods requires perfect alignment between the transmit and receive antennas. However, the harsh precondition of the perfect alignment in practical OAM communication system is difficult to satisfied all the time. To evaluate the effect of the misalignments on the array signal phases, we compare the helical phases of the single-mode OAM beam and multi-mode OAM beams in the perfect alignment and misalignment cases, as shown in Fig. \ref{Fig2}, in which the frequency is $f=2.36$ GHz, the radius of the transmit UCA is $R_{t} = 9\lambda$, and the transmission distance $D=300\lambda$. It can be seen from the figure that both the single-mode and multi-mode OAM beams are highly sensitive to the oblique angle of the receive UCA. Then, traditional detection methods that depends directly on circular phase gradient fail to work. Therefore, it is necessary to propose effective mode detection method for more general misalignment cases for OAM communication systems.

\begin{figure}[t] 
\setlength{\abovecaptionskip}{0cm}   
\setlength{\belowcaptionskip}{-0.2cm}   
\begin{center}
\includegraphics[scale=0.4]{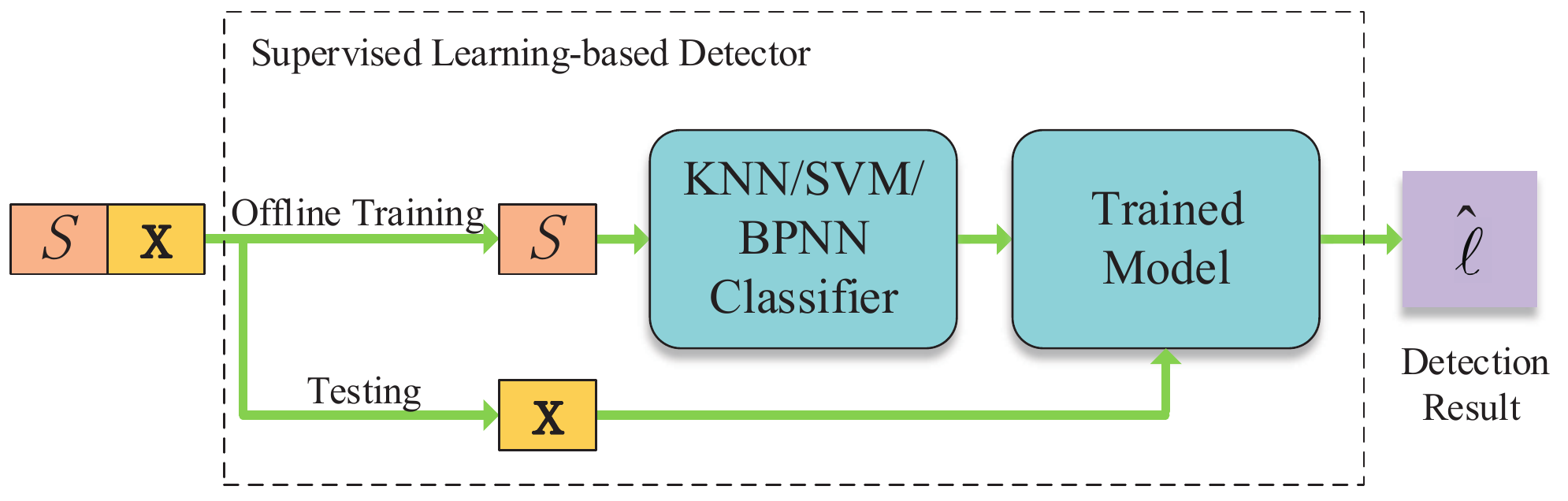}   
\end{center}
\caption{OAM mode detector model based on supervised learning.}
\label{Fig9}
\end{figure}
\subsection{OAM Mode Detection Based on Supervised Learning}
\begin{figure}[t]  
\setlength{\abovecaptionskip}{0cm}   
\setlength{\belowcaptionskip}{-0.2cm}   
\centering
\subfigure[]{\includegraphics[scale=0.55]{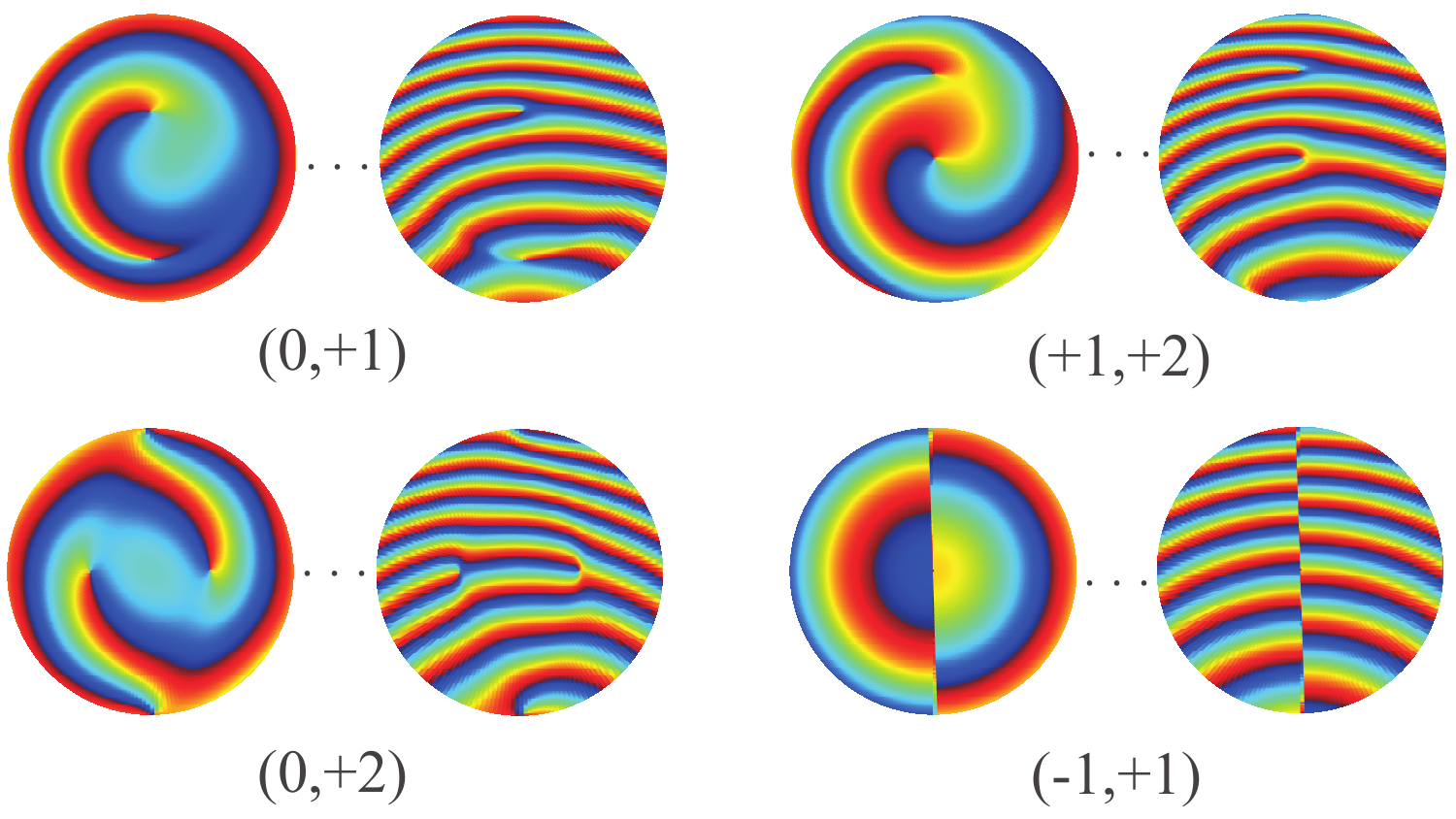}}
\quad
\subfigure[]{\includegraphics[scale=0.55]{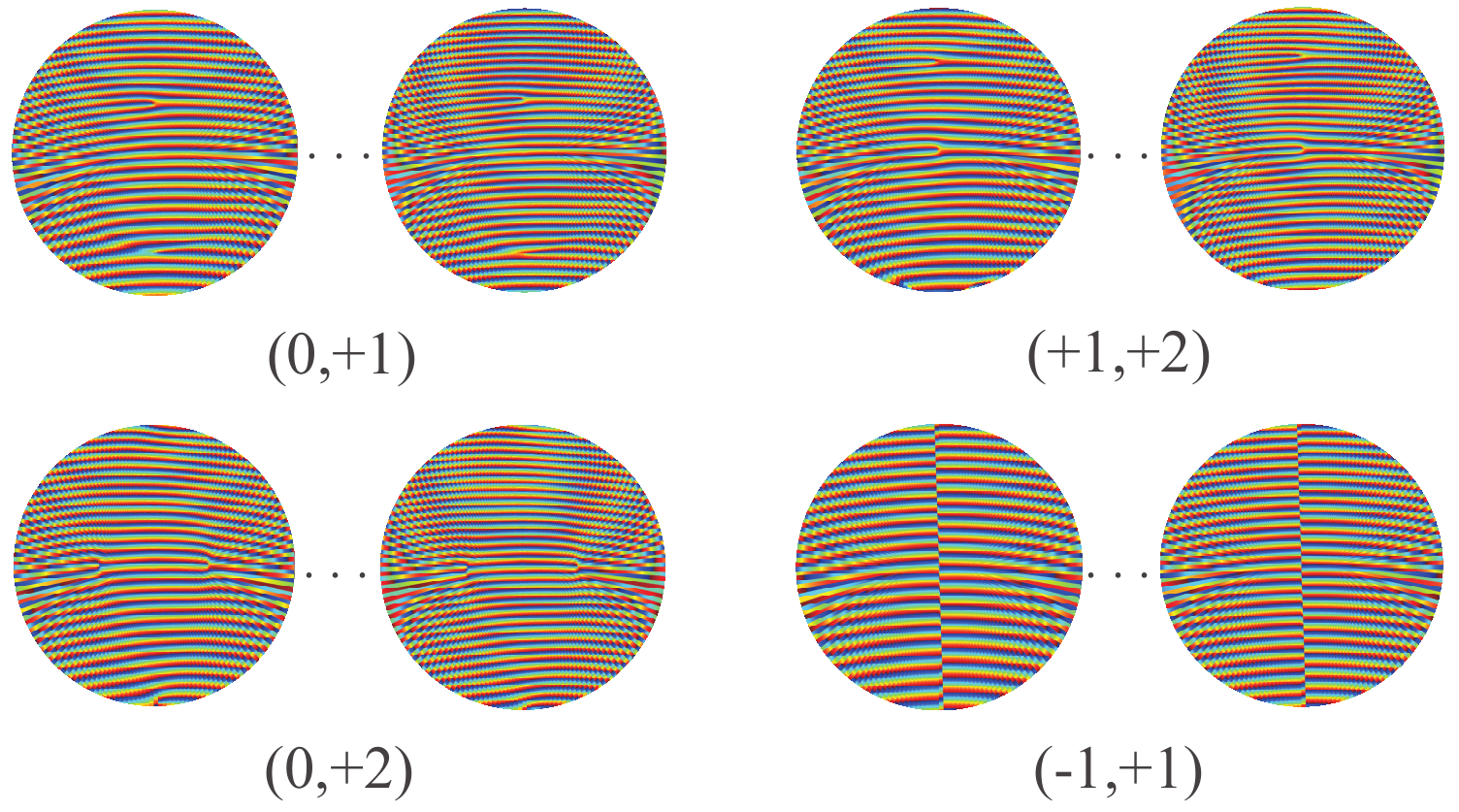}}
\caption{An example of the phase distributions of two superimposed OAM modes ${\left( {0,+1} \right),\left( {+ 1,+2} \right),\left( {0,+2} \right),\left( { - 1,+1} \right)}$: (a)${D}_{train}=300\lambda$, ${\alpha}_{train}=0 \sim 20^{\circ}$, (b)${D}_{test}=400\lambda$, ${\alpha}_{test}=20 \sim 30^{\circ}$.}
\label{Fig4}
\end{figure}

Three OAM mode detection methods based on supervised learning, i.e., KNN, SVM and BPNN classifiers, are proposed to detect which the single-mode is and which the multi-mode combination is. As shown in Fig. \ref{Fig9}, our OAM detection method first built different trained models for KNN, SVM and BPNN classifiers by training the phase sample data set $S$ in the offline training stage, from which the relationship between the OAM phase information and OAM mode can be obtained. Then, when the baseband signals are received in the testing stage,  the classifiers output the predicted OAM mode based on the corresponding train model and the received signals ${\bf{x}}$. Taking the dual-mode OAM as an example, Fig. \ref{Fig4} (a), (b) respectively show the phase distributions of two superimposed OAM modes in the offline training stage and the final testing stage to verify the generalization performance of the proposed method under different distances and oblique angles.

\subsubsection{OAM Mode Detection with KNN and SVM Classifiers}
In the existing classifiers, we first consider classifiers KNN and SVM which are mainly utilized for points/objects classification \cite{Machine2017Jiang}. For the KNN-based detection algorithm, the KNN classifier first obtain $K$ samples from $S$ most similar to the received signal ${\bf{x}}$ by comparing the distance between the phase of ${\bf{x}}$ and samples, and then determine the OAM mode of ${\bf{x}}$ according to the label of the $K$ samples $\{{\ell}^{p,q}_k|k=,1,2,\cdots,K\}$. Besides, for SVM-based detection algorithm, SVM first maps samples $S$ and received signal ${\bf{x}}$ to high-dimensional data set relying on non-linear mapping, and then searches for the optimal linear separation hyperplane on the high-dimensional data to detecte the OAM mode of ${\bf{x}}$.

\begin{figure}[t] 
\setlength{\abovecaptionskip}{0cm}   
\setlength{\belowcaptionskip}{-0.2cm}   
\begin{center}
\includegraphics[scale=0.35]{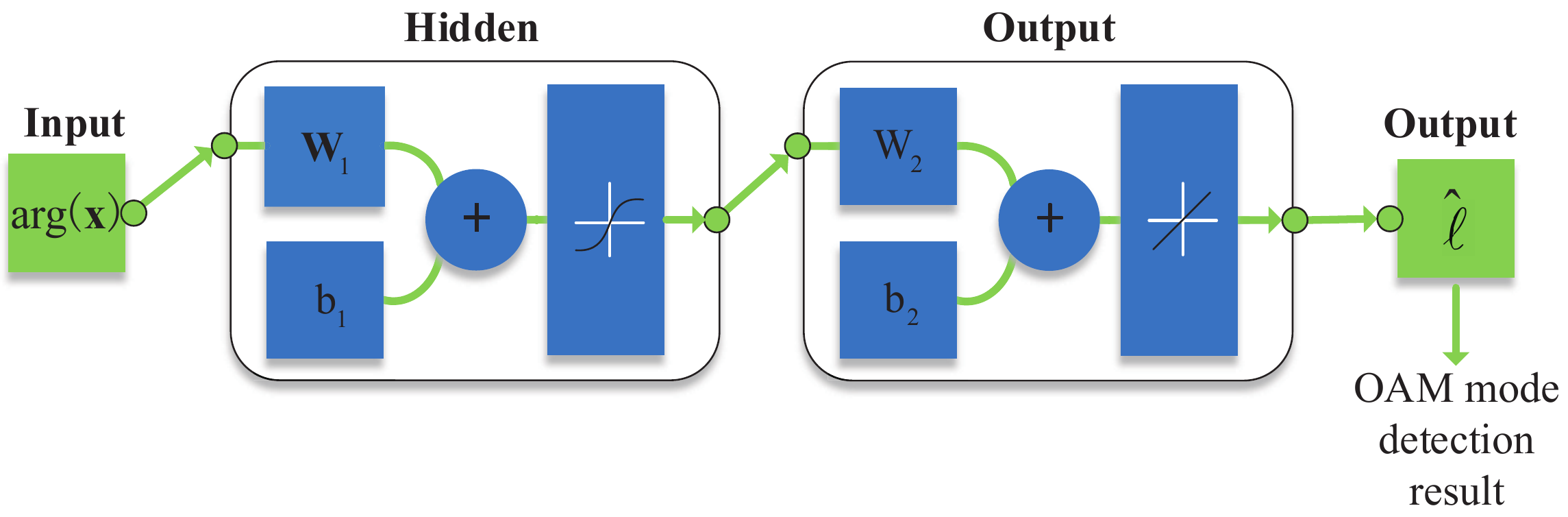}   
\end{center}
\caption{Two-layer BPNN structure applied to the OAM mode detection.}
\label{Fig5}
\end{figure}

\subsubsection{OAM Mode Detection with BPNN Classifier}
In this part, we mainly discuss the OAM mode detection method based on the two-layer BPNN. The neural network is generally composed of input layer, hidden layer and output layer. The system model of the proposed OAM mode detection method is shown in Fig. \ref{Fig5}, where $\bf W_1$, $\bf W_2$, $\bf b_1$ and $\bf b_2$ are the weights and bias of the network respectively. In the offline training stage, the sample data set $S$ is normalized and input in to the hidden layer of the network, and then is processed by the tanh nonlinear activation function
\begin{align}\label{Tanh}
{f_{\tanh }}(x) = \frac{{{e^x} - {e^{ - x}}}}{{{e^x} + {e^{ - x}}}},
\end{align}
and sent to the output layer. Thereafter, the output layer performs the inverse normalization and output the predicted label ${\hat{\ell}}^{p,q}$, which can be expressed as
\begin{align}\label{Output}
 {\hat{\ell}}^{p,q} = {f_{output}}({{\bf{W}}_2}{f_{\tanh }}({{\bf{W}}_1}{{\bf{\bar{x}}}(D_p,\alpha_q)} + {{\bf{b}}_1}) + {{\bf{b}}_2}).
\end{align}
Repeat the above progress and minimize the mean square error (MSE) loss function
\begin{equation*} \label{MSE}
{\rm{MSE}} = \frac{1}{\rm{PQ}}\sum\limits_{p = 1}^P\sum\limits_{q = 1}^Q ({\ell}^{p,q} - {\hat{\ell}}^{p,q})^2
\end{equation*}
by Bayesian regularization algorithm \cite{Song2013Hybrid}.
After that, the optimal weights and bias can be obtained by BPNN. In the testing stage, the phase of the received signal $\bf x$ is also normalized and input in to the network, and is processed similar to the offline training stage. Finally, the output layer output the detected OAM mode.

\section{Numerical Simulation and Results}
\begin{table}[t]  
\scriptsize
\centering
\caption{The results of single-mode OAM detection based on supervised learning.}
\label{Table1}
\begin{tabular}{ccccc}
\toprule
\multirow{2}{*}{Detection methods} & \multicolumn{2}{c}{OAM single-mode detection} & \multicolumn{2}{c}{OAM multi-mode detection}\\
\cmidrule(r){2-3} \cmidrule(r){4-5}
&  Accuracy      &  Testing time
&  Accuracy      &  Testing time \\
\midrule
KNN              & 91.16{\rm{\% }}     &0.1720s      & 74.90{\rm{\% }}     & 0.1250s \\
SVM              & 95.75{\rm{\% }}     &0.1440s      & 98.56{\rm{\% }}     & 0.0400\\
BPNN   &100.00{\rm{\% }}     &0.0020s      &100.00{\rm{\% }}     &0.0020s \\
\bottomrule
\end{tabular}
\end{table}
\begin{figure}[t]  
\setlength{\abovecaptionskip}{0cm}   
\setlength{\belowcaptionskip}{-0.2cm}   
\centering
\subfigure[]{\includegraphics[height=4.2cm]{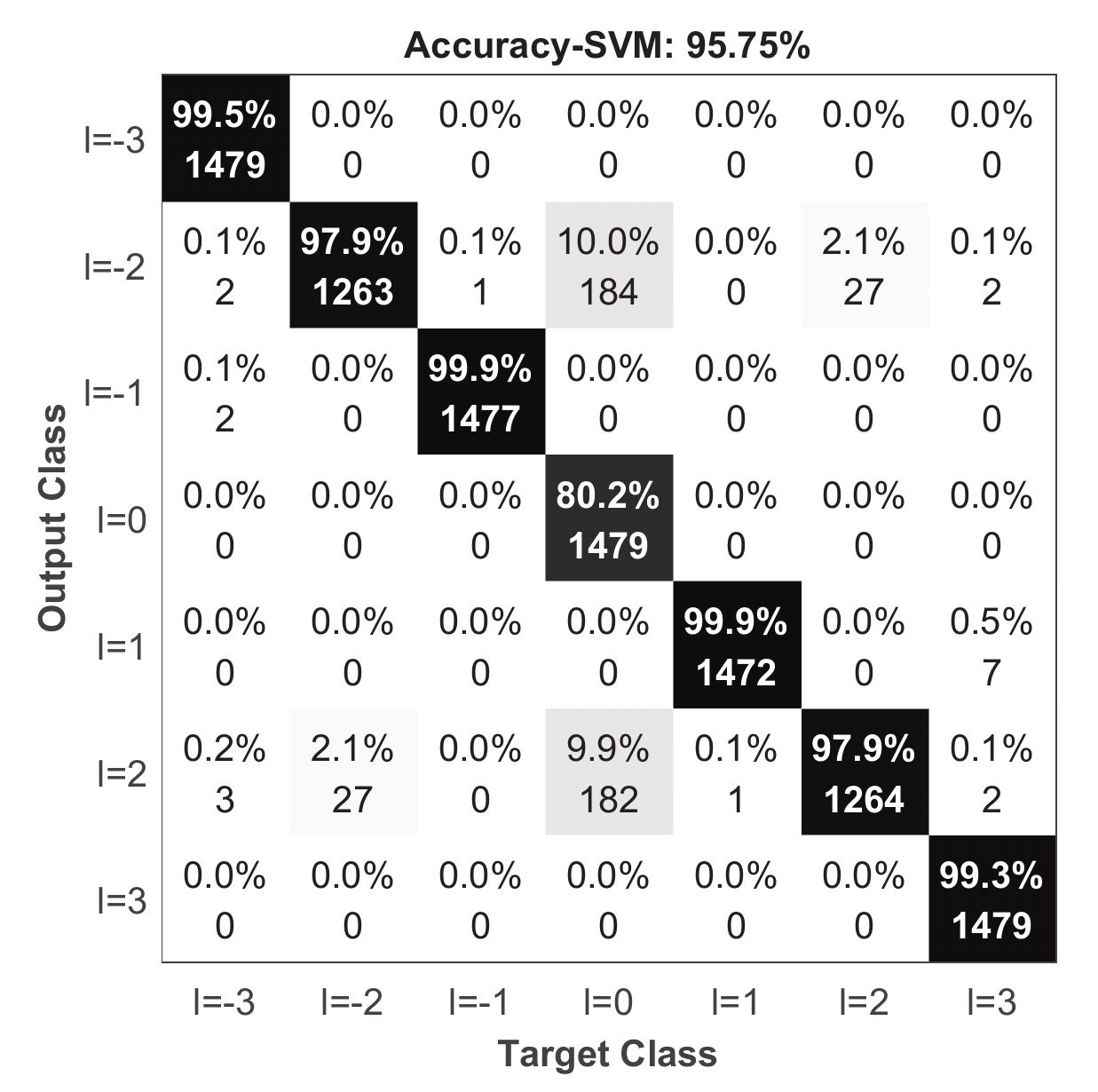}}\hspace{-0.8mm}
\subfigure[]{\includegraphics[height=4.2cm]{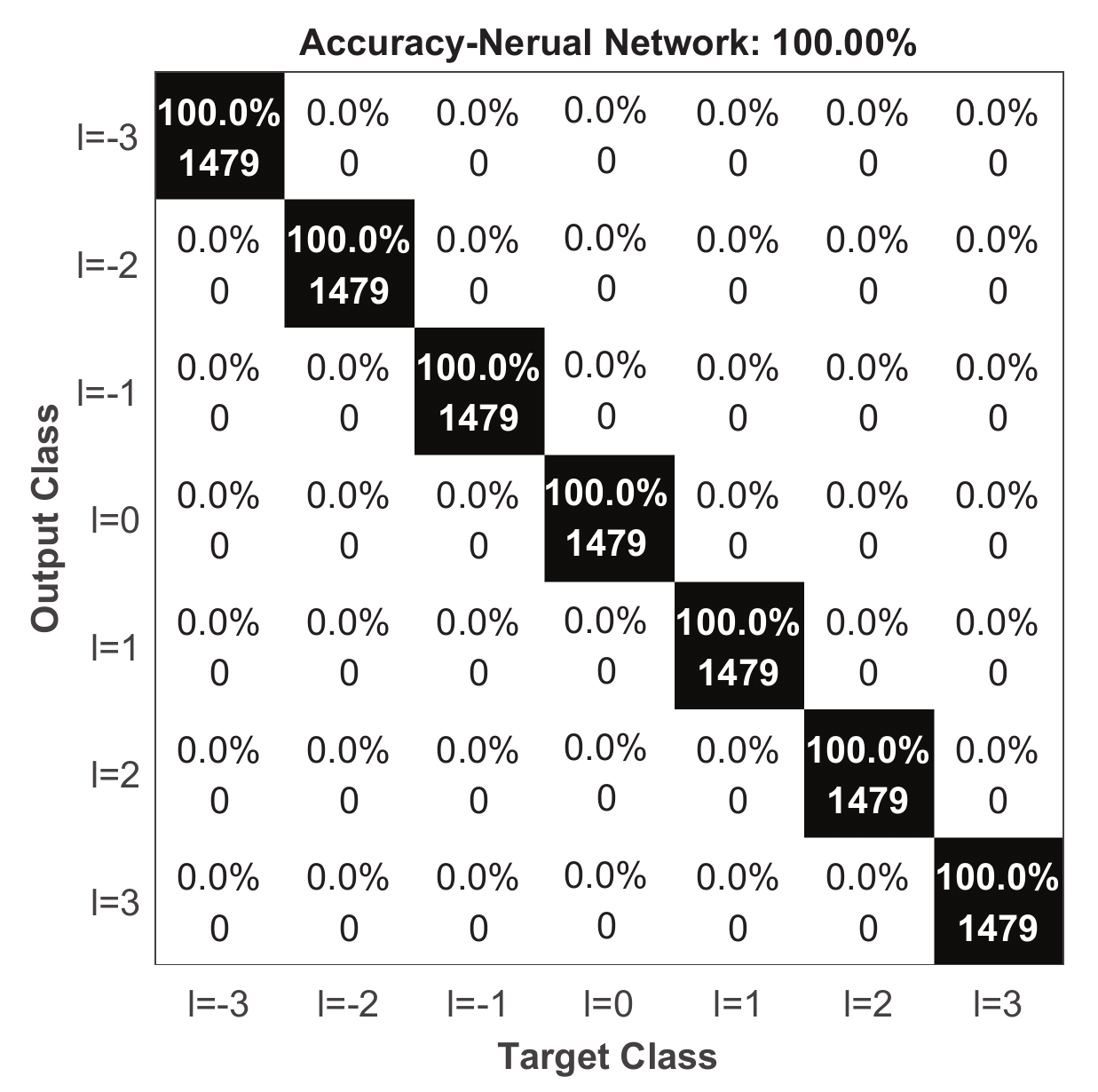}}
\caption{The results of the single-mode OAM detection: (a)the confusion matrix of SVM, (b)the confusion matrix of BPNN.}
\label{Fig6}
\end{figure}
\begin{figure}[t]  
\setlength{\abovecaptionskip}{0cm}   
\setlength{\belowcaptionskip}{-0.2cm}   
\centering
\subfigure[]{
\includegraphics[height=4.2cm]{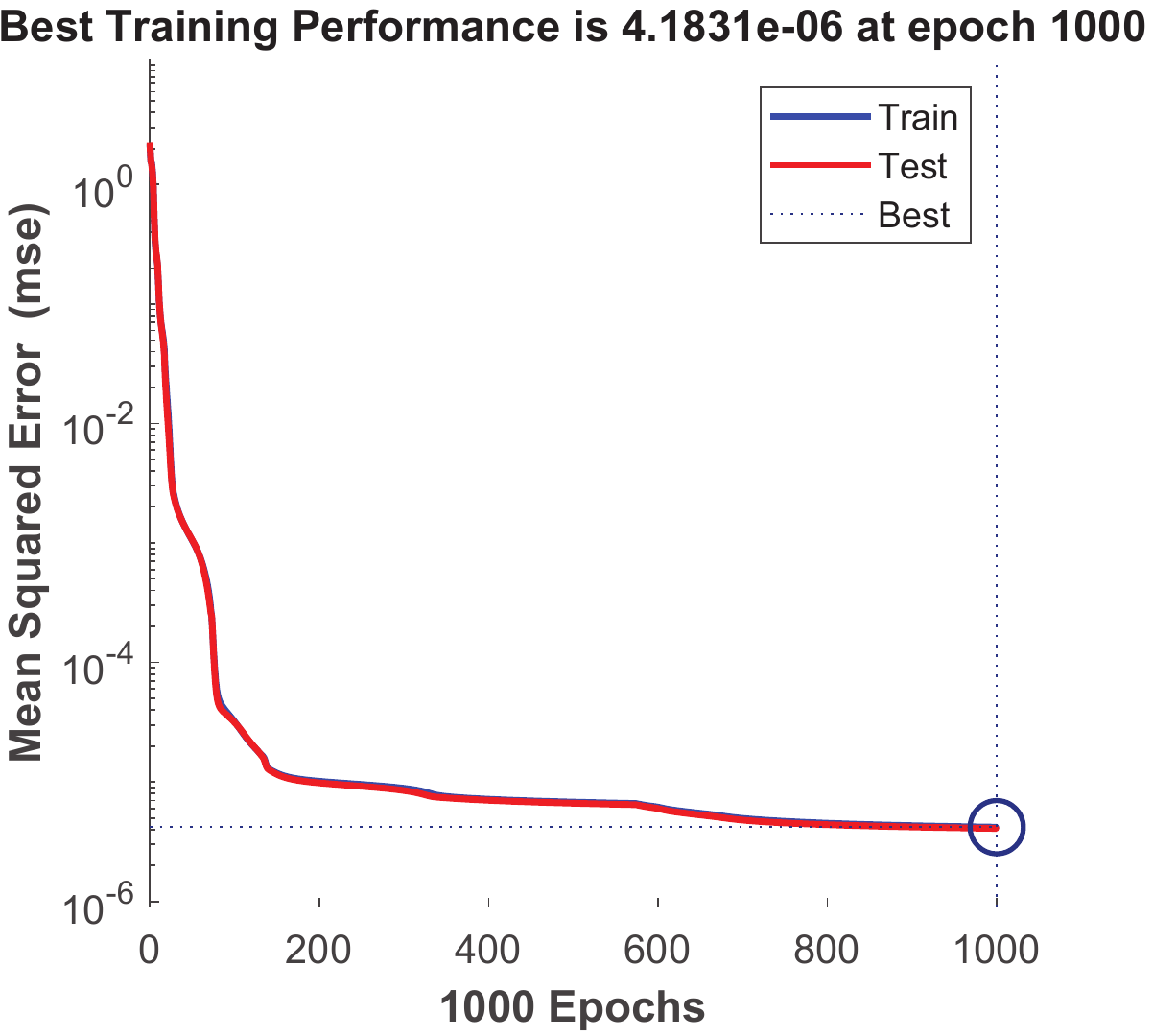}}\hspace{-6mm}
\subfigure[]{
\includegraphics[height=4.2cm]{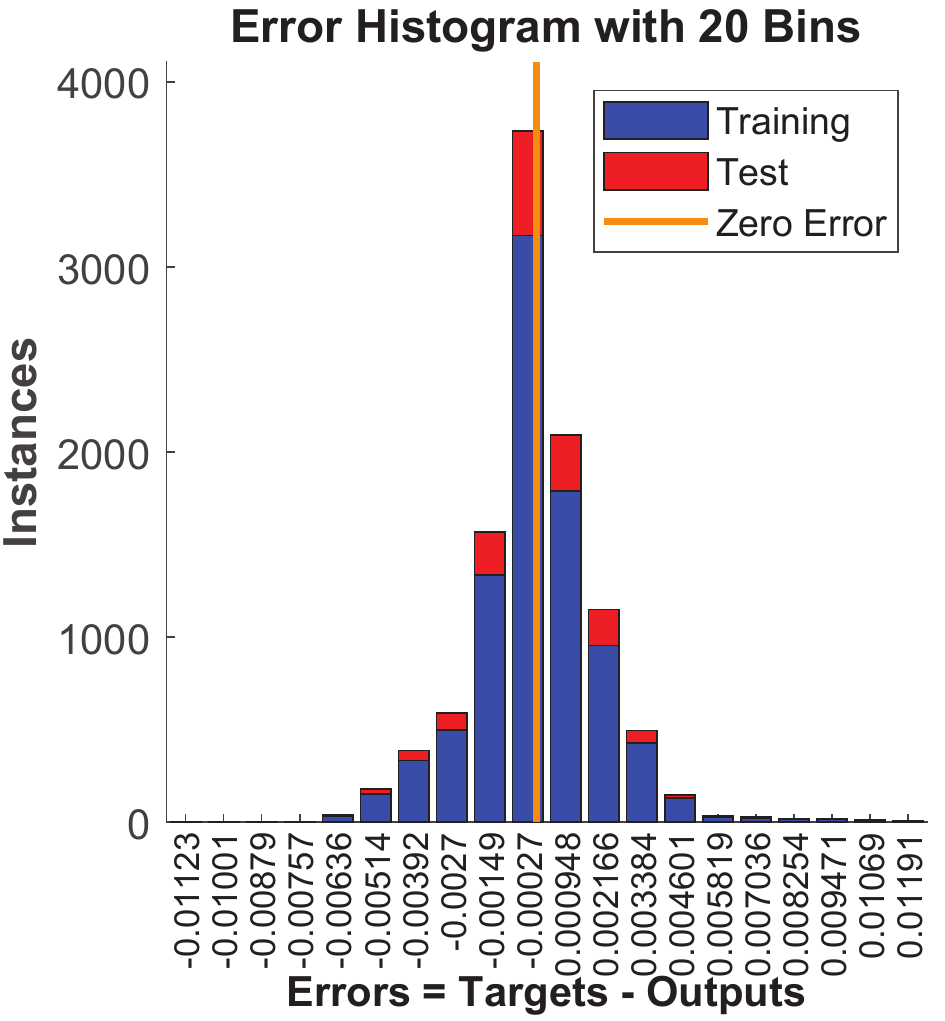}
}

\subfigure[]{
\includegraphics[scale=0.47]{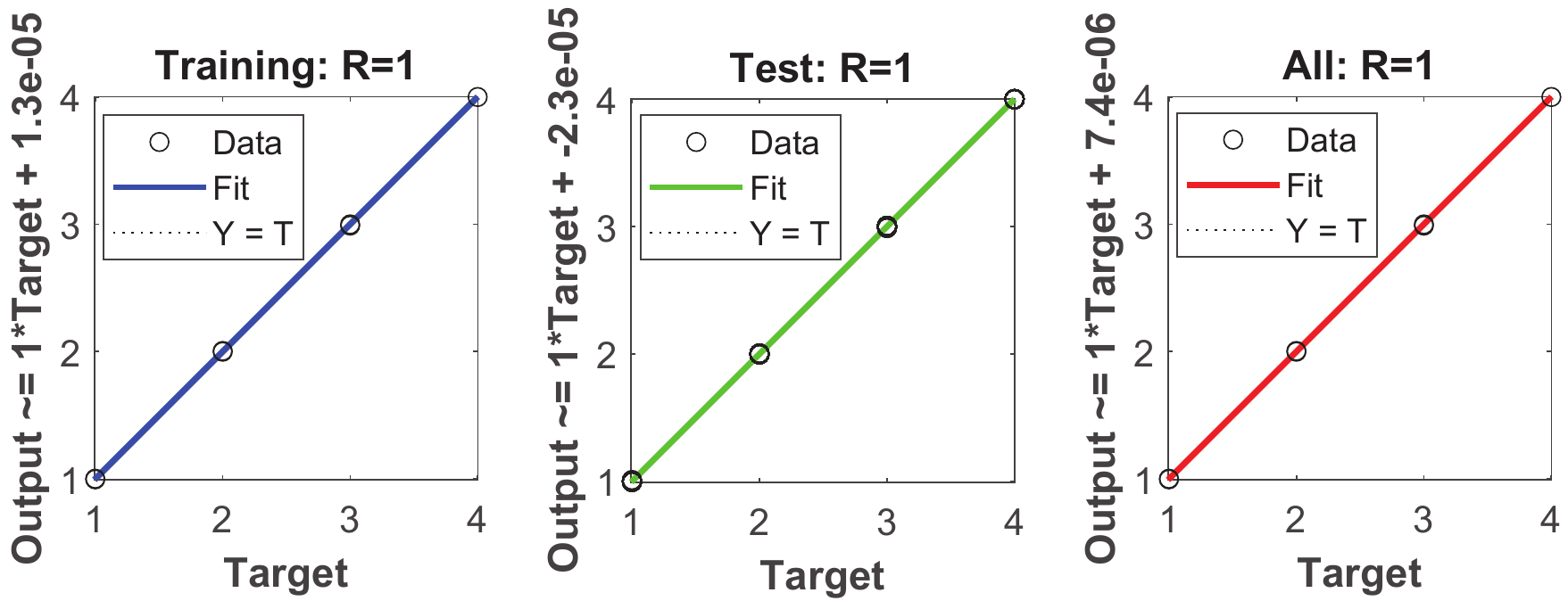}
}
\caption{
The results of multi-mode OAM detection based on BPNN: (a)training MSE performance, (b)training error histogram, (c)training regression.}
\label{Fig7}
\end{figure}
\begin{figure}[t]  
\setlength{\abovecaptionskip}{0cm}   
\setlength{\belowcaptionskip}{-0.2cm}   
\centering
\subfigure[]{
\includegraphics[height=4.2cm]{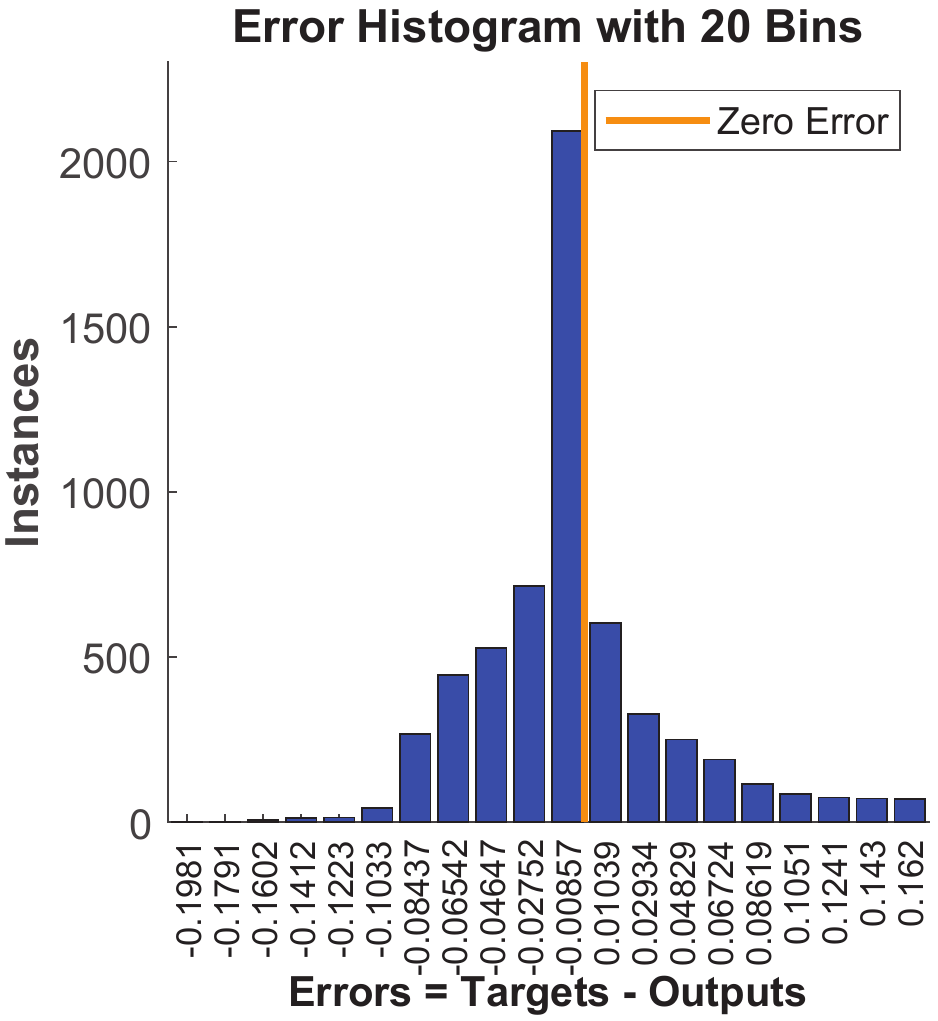}     
}
\subfigure[]{
\includegraphics[height=4.2cm]{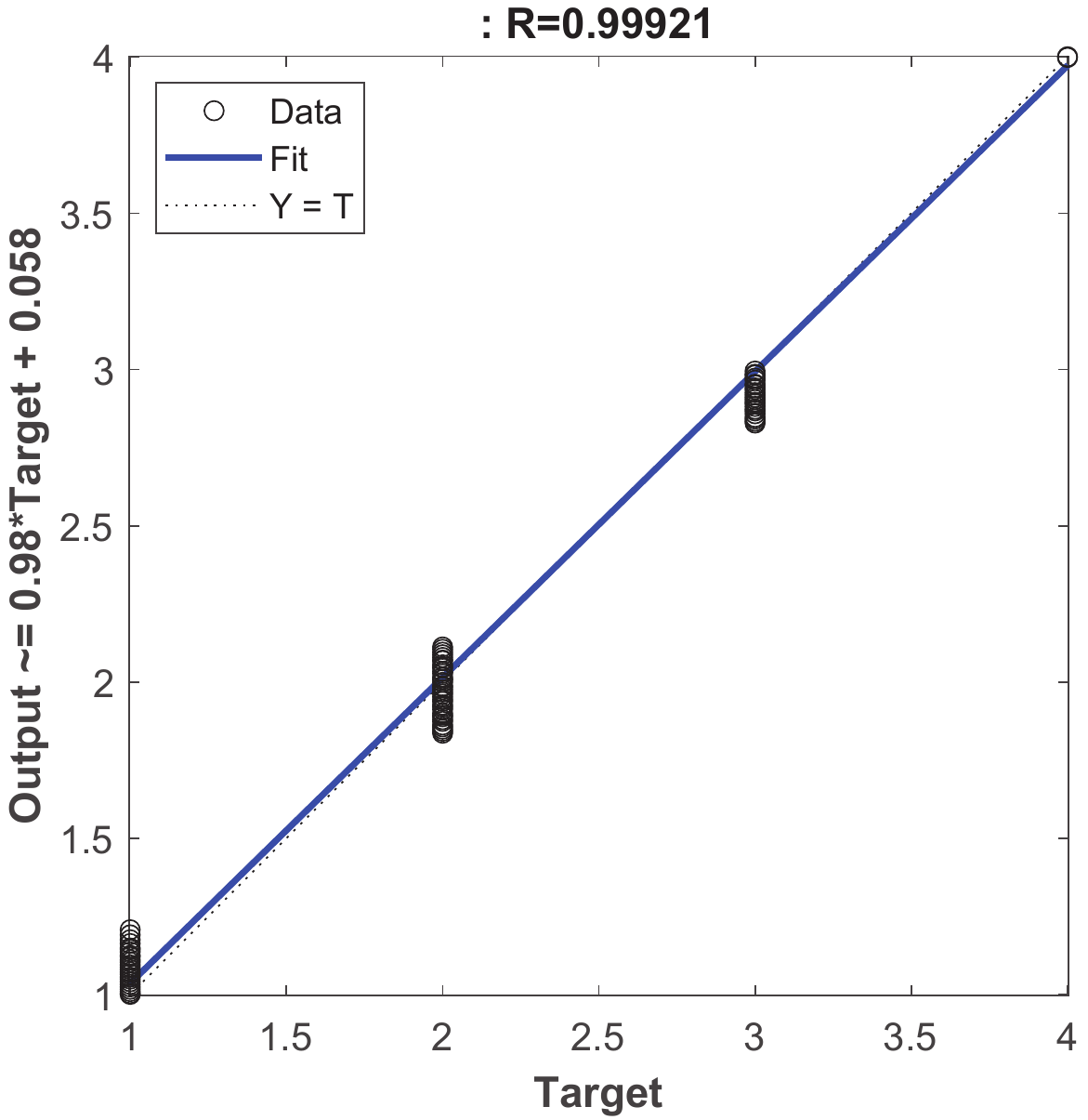}  
}
\caption{
The results of multi-mode OAM detection based on BPNN: (a)testing error histogram, (b)testing regression.}
\label{Fig8}
\end{figure}

In the single-mode OAM detection, we assume $N_t=8$, $N_r=10$, $U=7$ with $\ell_u=\pm3,\pm2,\pm1,0$, $K=1$, and the neural network having 10 neurons in the hidden layer, and choose $P=25$ with $D_p=150\lambda, 158\lambda, \cdots, 350\lambda$, $Q=100$ with ${\alpha}_{q}=0^{\circ}, 0.2^{\circ}, \cdots, 20^{\circ}$. In the multi-mode OAM  detection, we assume $N_t=6$, $N_r=10$, $U=4$ with the mixed OAM mode $(\ell_1, \ell_2) \in \left\{(0,+1),(+1,+2),(0,+2),(-1,+1)\right\}$. Other parameters are the same as the single-mode OAM detection. In the testing stage, we assume the receiver is randomly located at $\{(D,\alpha)|360\lambda\leq D\leq560\lambda, 20^{\circ}\leq \alpha\leq30^{\circ}\}$.

Table \ref{Table1} shows the detection accuracy and testing time of three different OAM detection methods. It can be seen from the table that  all the three methods are robust in the non-parallel misalignment case. Besides, compared with the traditional KNN-based and SVM-based detection methods, the OAM detection method based on the two-layer BPNN has the best detection accuracy and the least testing time. Fig. \ref{Fig6} compares the single-mode OAM detection results of the SVM-based method and the two-layer BPNN method. It can be seen from the figure that the generalization performance of the two-layer BPNN method is better than SVM-based method.
Fig. \ref{Fig7}(a) shows the relationship between epochs and MSE of the multi-mode OAM detection based on the two-layer BPNN in the offline training stage. Fig. \ref{Fig7}(b) shows the error distribution of the sample data set, where zero error means no error. Fig. \ref{Fig7}(c) is the training regression of the multi-mode OAM detection based on two-layer BPNN in the offline training stage. Based on the the optimal weights and bias obtained by Fig. \ref{Fig7}, Fig. \ref{Fig8} (a) and Fig. \ref{Fig8} (b) show the generalization performance of the multi-mode OAM detection based on the two-layer BPNN. It can be seen from Fig. \ref{Fig8} (b) that the slope of fitting curve is closed to $1$, which verifies the excellent multi-mode detection performance of the proposed method. It is necessary to explain that the abscissa $\{1,2,3,4\}$ in Fig. \ref{Fig7}(c) and Fig. \ref{Fig8}(b) respectively represent four OAM multi-mode combination $\{{(\ell_1, \ell_2)} \in \left\{ {\left( {0,+1} \right),\left( {+1,+2} \right),\left( {0,+2} \right),\left( {-1,+1} \right)} \right\}\}$.

\section{Conclusions}
In this paper, we first show the OAM phase structures are sensitive to the misalignment between the transmit and receive antennas, which results in classical phase gradient-based OAM mode detection methods fail to work. Motivated by this fact, we propose three OAM detection methods based on supervised learning for the UCA-based OAM communication system in more general alignment or non-parallel misalignment cases, which can realize single-mode and multi-mode OAM detection by the KNN, SVM and BPNN classifiers, respectively. The simulation results show that all the three methods are robust in the non-parallel misalignment case, and compared with KNN and SVM, the two-layer BPNN not only has good performance in the offline training stage, but also shows the best generalization performance in the testing stage.

\section*{Acknowledgment}
This work was supported in part by Natural Science Basic Research Program of Shaanxi (Program No. 2021JZ-18), Natural Science Foundation of Guangdong Province of China under Grant 2021A1515010812, the open research fund of National Mobile Communications Research Laboratory, Southeast University under Grant number 2021D04, the Fundamental Research Funds for the Central Universities, and the Innovation Fund of Xidian University.

\bibliographystyle{IEEEtran}
\bibliography{IEEEabrv,OAM-ML}
\end{document}